\documentclass[a4paper,11pt]{article}

\usepackage{jheppub} 

\usepackage[T1]{fontenc} 
\usepackage{slashed}

\newcommand{\met}{\not\!\!\!E_{T}}
\newcommand{\bea} {\begin{eqnarray}}
\newcommand{\eea} {\end{eqnarray}}

\title{\boldmath Little Higgs Dark Matter after PandaX-II/LUX-2016 and LHC Run-1}

\author[a,b]{Lei Wu,}
\author[c]{Bingfang Yang,}
\author[d]{Mengchao Zhang,}

\affiliation[a]{Department of Physics and Institute of Theoretical Physics,\\ Nanjing Normal University, Nanjing, Jiangsu 210023, China}
\affiliation[b]{ARC Centre of Excellence for Particle Physics at the Terascale, School of Physics, \\ The University of Sydney, NSW 2006, Australia}
\affiliation[c]{Institution of Theoretical Physics,\\ Henan Normal University, Xinxiang 453007, China}
\affiliation[d]{Center for Theoretical Physics and Universe, Institute for Basic Science (IBS),
Daejeon 34051, Korea}

\emailAdd{leiwu1@sydney.edu.au}
\emailAdd{yangbingfang@htu.edu.cn}
\emailAdd{mczhang@itp.ac.cn}

\abstract{In the Littlest Higgs model with $T$ parity (LHT), the $T$-odd heavy photon ($A_H$) is weakly interacting and can play the role of dark matter. We investigate the lower limit on the mass of $A_H$ dark matter under the constraints from Higgs data, EWPOs, $R_b$, Planck 2015 dark matter relic abundance, PandaX-II/LUX 2016 direct detections and LHC-8 TeV monojet results. We find that (1) Higgs data, EWPOs and $R_b$ can exclude the mass of $A_H$ up to 99 GeV. To produce the correct dark matter relic abundance, $A_H$ has to co-annihilate with $T$-odd quarks ($q_H$) or leptons ($\ell_H$); (2) the LUX (PandaX-II) 2016 data can further exclude $m_{A_H}<380 (270)$ GeV for $\ell_H$-$A_H$ co-annihilation and $m_{A_H}<350 (240)$ GeV for $q_H-A_H$ co-annihilation; (3) LHC-8 TeV monojet result can give a strong lower limit, $m_{A_H}>540$ GeV, for $q_H$-$A_H$ co-annihilation; (4) future XENON1T (2017) experiment can fully cover the parameter space of $\ell_H$-$A_H$ co-annihilation and will push the lower limit of $m_{A_H}$ up to about 640 GeV for $q_H$-$A_H$ co-annihilation.}

\begin{document}
\maketitle
\flushbottom

\section{Introduction}
\label{sec:intro}
The discovery of 125 GeV Higgs boson \cite{higgs-atlas,higgs-cms} is a great step towards elucidating the electroweak symmetry breaking mechanism. However, without protection by a symmetry, the Standard Model (SM) Higgs boson mass should be quadratically sensitive to the cutoff scale $\Lambda$ (usually refers to Planck scale $\sim 10^{19}$ GeV) via radiative corrections, which renders the theory with $m_h \ll \Lambda$ rather unnatural. Besides, the SM cannot explain other convincing observations, such as the dark matter abundance in the Universe. In fact, the electroweak naturalness problem and dark matter are widely considered as major motivations for new physics beyond the SM.

Among various extensions of the SM, the Littlest Higgs model with $T$-parity (LHT) \cite{lht-1,lht-2,lht-3} is one of the most promising candidates. It can not only successfully solve the electroweak naturalness problem but also provide a viable dark matter candidate. On the theoretical side, the LHT model is based on a non-linear $\sigma$ model describing an $SU(5)/SO(5)$ symmetry breaking, which extends the Littlest Higgs model \cite{lh-1,lh-2,lh-3} by introducing the discrete symmetry $T$ parity. All of the global symmetries that protect the Higgs mass are explicitly broken. The Higgs boson is realized as a pseudo-Nambu-Goldstone boson of the broken global symmetry. With the collective symmetry breaking mechanism, all quadratically divergent contributions to the Higgs mass only first appear at two-loop level, and thus are sufficiently small. On the phenomenological side, the introduction of $T$-parity in the LHT model can relax the electroweak precision observables (EWPOs) bound on the breaking scale $f$ by preventing the tree-level contributions from the heavy gauge bosons \cite{lh-ewpos-1,lh-ewpos-2,lh-ewpos-3,lh-ewpos-4,lh-ewpos-5,lh-ewpos-6,lht-ewpos-1} and lead to an abundant phenomenology at the LHC \cite{lht-phe-1,lht-phe-2,lht-phe-3,lht-phe-4,lht-phe-4.5,lht-phe-5,lht-phe-5.5,lht-phe-6,lht-phe-7,lht-phe-8,lht-phe-9,lht-phe-10,lht-phe-11,lht-phe-12,lht-phe-13,lht-phe-14,lht-phe-15,lht-phe-16,lht-phe-16.5,lht-phe-17,lht-phe-18,lht-phe-19}. Besides, the $T$-parity guarantees that the lightest $T$-odd particle (LTP) is stable so that it can naturally serve as the dark matter candidate if it is charge-neutral and colorless. One of such candidates is $T$-odd partner of the hypercharge gauge boson $A_H$ \footnote{Besides $A_H$, $T$-odd partner of neutrino $\nu_H$ can be a dark matter candidate as well. However, the direct detection experiments have excluded this possibility because the cross section of elastic scattering between $\nu_H$ and nucleus is about $4\sim5$ order of magnitude larger than the current experimental bound \cite{Servant:2002hb}.}.

The phenomenology of heavy photon dark matter has been studied in \cite{lht-dm,lht-dm-wanglei,lht-dm-chuanren}. In general, there are two ways to achieve the correct dark matter relic abundance. One is that two $A_H$ dark matter annihilate into SM particles, which is mainly through the $s$-channel via exchanging the Higgs boson. However, due to the constraints of Higgs data and EWPOs, the mass of heavy photon is heavier than $m_h/2$ \cite{lht-fit-tonini-1,lht-fit-tonini-2,lht-fit-yang-1,lht-fit-yang-2,lht-fit-wang}. Thus, without resonant enhancement, the pair annihilation cross section of $A_H$ is usually too small to satisfy the observed dark matter relic density. The other is that the $A_H$ dark matter co-annihilates with other $T$-odd particles, such as mirror quarks $q_H$ or leptons $\ell_H$. The co-annihilation of dark matter in simplified models has been studied in \cite{Baker:2015qna}.

In this work, we will investigate the lower bound on the mass of $A_H$ dark matter co-annihilations in the LHT model. We will consider various relevant constraints, including Higgs data, EWPOs, $R_b$, Planck dark matter relic abundance, PandaX-II/LUX-2016 results and LHC-8 TeV monojet result. This paper is organized as follows. In section \ref{section2}, we give a brief description of the heavy photon dark matter and $T$-odd fermion sector of the LHT model. In section \ref{section3}, we examine various constraints on $A_H$ dark matter. Finally, we draw our conclusions in section \ref{section4}.

\section{Littlest Higgs model with $T$-parity}\label{section2}
\subsection{Heavy Photon}
The LHT model is a realization of non-linear $\sigma$ model, which is based on the coset space $SU(5)/SO(5)$. The global symmetry $SU(5)$ is spontaneously broken into $SO(5)$ at TeV scale by the vacuum expectation value (VEV) of the $\Sigma$ field,
\begin{eqnarray}
\Sigma_0=\langle\Sigma\rangle
\begin{pmatrix}
{\bf 0}_{2\times2} & 0 & {\bf 1}_{2\times2} \\
0 & 1 &0 \\
{\bf 1}_{2\times2} & 0 & {\bf 0}_{2\times 2}
\end{pmatrix}.
\end{eqnarray}
In the meantime, the VEV $\Sigma_0$ breaks the gauged subgroup $\left[ SU_1(2) \times U_1(1) \right] \times \left[ SU_2(2) \times U_2(1) \right]$ of $SU(5)$ down to the diagonal SM electroweak gauge group $SU_L(2) \times U_Y(1)$. In the end, there are 4 new heavy gauge bosons $W_{H}^{\pm},Z_{H},A_{H}$, whose masses are given at $\mathcal O(v^{2}/f^{2})$ by
\begin {equation}
M_{W_{H}}=M_{Z_{H}}=gf(1-\frac{v^{2}}{8f^{2}}),~~M_{A_{H}}=\frac{g'f}{\sqrt{5}}
(1-\frac{5v^{2}}{8f^{2}})\label{vmass}
\end {equation}
where $g$ and $g'$ are the SM $SU_L(2)$ and $U_Y(1)$ gauge couplings, respectively. In order to match the SM prediction for the gauge boson masses, the VEV $v$ needs to be redefined via the functional form
\begin{equation}
v = \frac{f}{\sqrt{2}} \arccos{\left( 1 - \frac{v_\textrm{SM}^2}{f^2} \right)} \simeq v_\textrm{SM} \left( 1 + \frac{1}{12} \frac{v_\textrm{SM}^2}{f^2} \right) ,
\end{equation}
where $v_{SM}$ = 246 GeV. The heavy photon $A_{H}$ is typically the lightest $T$-odd state and thus can be a possible candidate for dark matter. The only direct coupling of a pair of $A_H$ to the SM sector is via the Higgs boson, resulting in weak-strength cross sections for $A_H$ scattering into SM states.

\subsection{$T$-odd Fermions}
Two fermion $SU(2)$ doublets $q_1$ and $q_2$ are introduced in the LHT model, where $q_i~(i = 1, 2)$ is transformed as a doublet under $SU(2)_i$, and $T$-parity interchanges these two doublets. The $T$-even combination of these two doublets is considered as the SM $SU(2)$ doublet, while the $T$-odd combination has to gain a TeV scale mass. The fermion $SU(2)$ doublets $q_1$ and $q_2$ are embedded into incomplete $SU(5)$ multiplets
$\Psi_1$ and $\Psi_2$ as $\Psi_1 = (q_1, 0, 0_2)^T$ and $\Psi_2 = (0_2, 0, q_2)^T$, in which $0_2 = (0, 0)^T$. Besides, in order to generate masses to the heavy fermions, a $SO(5)$ multiplet $\Psi_c=(q_c,\chi_c,\tilde{q}_c)^T$ is introduced as well. The transformation of $\Psi_c$ under the $SU(5)$ is non-linear: $\Psi_c \to U\Psi_c$, where $U$ is the unbroken $SO(5)$ rotation and is a non-linear representation of the $SU(5)$. Then, the $T$-invariant Lagrangian for the mass terms of the $T$-odd fermions can be written as follows:
\begin{eqnarray}
{\cal L}_\kappa &=& -\kappa f (\bar{\Psi}_2\xi\Psi_c + \bar{\Psi}_1\Sigma_0\Omega\xi^\dagger\Omega\Psi_c) + ~{\rm h.c.}
\end{eqnarray}
Here $\Omega=diag(1, 1, -1, 1, 1)$. It should be noted that the non-linear field $\xi$ contains the Higgs field, which can generate the masses of the $T$-odd quarks up to ${\cal O}(v^2/f^2)$ as,
\begin{eqnarray}
m_{d^i_H}=\sqrt{2}\kappa_{d_i} f, \quad m_{u^i_H}=\sqrt{2}\kappa_{u_i} f(1-\frac{v^2}{8f^2})\label{fmass1}
\end{eqnarray}
where $\kappa_{q_i}(q=u,d)$ are the diagonalized Yukawa couplings of the $T$-odd quarks. Similarly, the masses of the $T$-odd leptons are given by,
\begin{eqnarray}
m_{\ell^i_H}=\sqrt{2}\kappa_{\ell_i} f, \quad m_{\nu^i_H}=\sqrt{2}\kappa_{\nu_i} f(1-\frac{v^2}{8f^2})\label{fmass2}
\end{eqnarray}
where $\kappa_{\ell_i}$ and $\kappa_{\nu_i}$ are the diagonalized Yukawa couplings of $T$-odd leptons and neutrinos, respectively. From Eqs.~(\ref{vmass},\ref{fmass1},\ref{fmass2}), we note that only if $\kappa_{q_i,\ell_i,\nu_i} \gtrsim 0.11$, the heavy photon $A_H$ can become the LTP for a given value of $f$. For simplicity, we assume the universal Yukawa couplings $\kappa_{\ell_i}=\kappa_{\nu_i}=\kappa_\ell$ and $\kappa_{u_i}=\kappa_{d_i}=\kappa_q$, and require the Yukawa couplings $\kappa_{\ell,q}>0.11$.

\section{Constraints on Heavy Photon $A_H$ Dark Matter}\label{section3}
\subsection{Higgs data, EWPO and $R_b$}
In the LHT model, the nature of composite Higgs leads to the deviation of the Higgs gauge couplings from the SM values at the order of $v^2/f^2$. Moreover, mixing of the SM top with the $T$-even top partner ($T^+$) induces shifts in the Higgs couplings to gluons and photons. Here we list the relevant tree-level Higgs couplings for our fitting,
\bea
\begin{array}{ll}
hW^+W^-:~~\frac{2m_W^2}{v}(1-\frac{1}{6}\frac{v^2}{f^2})g^{\mu\nu},&
hZZ:~~\frac{2m_Z^2}{v}(1-\frac{1}{6}\frac{v^2}{f^2})g^{\mu\nu},\\
ht\bar{t}:~~-\frac{m_t}{v}\left[1+\frac{v^2}{f^2}(-\frac{2}{3}+\frac{R^2}{(1+R^2)^2})\right],&
hT^+\bar{T}^-:~~\frac{m_T}{v}\frac{R^2}{(1+R^2)^2}\frac{v^2}{f^2},\\
\end{array}
\label{coupling}
\eea
where $R$ is the mixing angle between the top quark and $T^+$ quark. The loop-induced couplings $hgg$ and $h\gamma\gamma$ are given in \cite{lht-phe-3}. Besides, there are two possible ways to construct $T$-invariant Lagrangians of the Yukawa interactions of the charged leptons and down-type quarks. Up to $\mathcal{O}\left( v_{SM}^4/f^4 \right)$, the ratios of the down-type quark Yukawa couplings $g_{hd\bar{d}}$ with respect to the SM prediction are given by \cite{lht-phe-3},
\begin{eqnarray}
    \frac{g_{h \bar{d} d}}{g_{h \bar{d} d}^{SM}} &=& 1-
        \frac{1}{4} \frac{v_{SM}^{2}}{f^{2}} + \frac{7}{32}
        \frac{v_{SM}^{4}}{f^{4}} \qquad \text{Case A} \nonumber \\
    \frac{g_{h \bar{d} d}}{g_{h \bar{d} d}^{SM}} &=& 1-
        \frac{5}{4} \frac{v_{SM}^{2}}{f^{2}} - \frac{17}{32}
        \frac{v_{SM}^{4}}{f^{4}} \qquad \text{Case B}.
    \label{hdd}
\end{eqnarray}
In our following fitting, we will confront the above modified Higgs couplings $hVV$, $hf\bar{f}$, $hgg$ and $h\gamma\gamma$ with the Higgs data for both cases.

In the LHT model, the electroweak precision observables $S$ and $T$ are changed by the modified Higgs gauge couplings $hVV$ \cite{lht-ewpos-1}. Furthermore, the top partner can correct the propagators of the electroweak gauge bosons at one-loop level. The UV operators can also contribute to the $S$ and $T$ parameters \cite{zyhan}. We take the couplings of the UV operators as unity \cite{lht-fit-tonini-1}. Besides, the new mirror fermions and new gauge bosons can contribute to the $Zb\bar{b}$ coupling at one-loop level \cite{lht-rb-1,lht-rb-2,lht-rb-3}. We will include the EWPOs and $R_b$ constraints in our study.

On the other hand, the current LHC direct searches for the multi-jet with the transverse missing energy can also produce the bounds on the parameter space of the LHT model. However, they are not strong enough to push the exclusion limits much beyond the indirect constraints \cite{lht-fit-tonini-1}. In particular, the ATLAS and CMS collaborations performed the searches for the vector-like top partner in different final states $bW$, $tZ$ and $th$. During the LHC Run-1, they excluded the masses of the top partners up to about 700 GeV \cite{tp-atlas,tp-cms}. However, those bounds depend on the assumptions of the group representations of top partners and their decay channels. In addition to the conventional decay channels ($bW$, $tZ$ and $th$), the $T$-even top partner $T^+$ can decay to $T^-A_H$, which can weaken the current LHC bounds on top partner in the LHT model \cite{lht-phe-18}. So in our scan, we consider the indirect constraints including Higgs data, EWPOs and $R_b$.

\begin{figure}[ht]
\centering
\includegraphics[width=7cm,height=8cm]{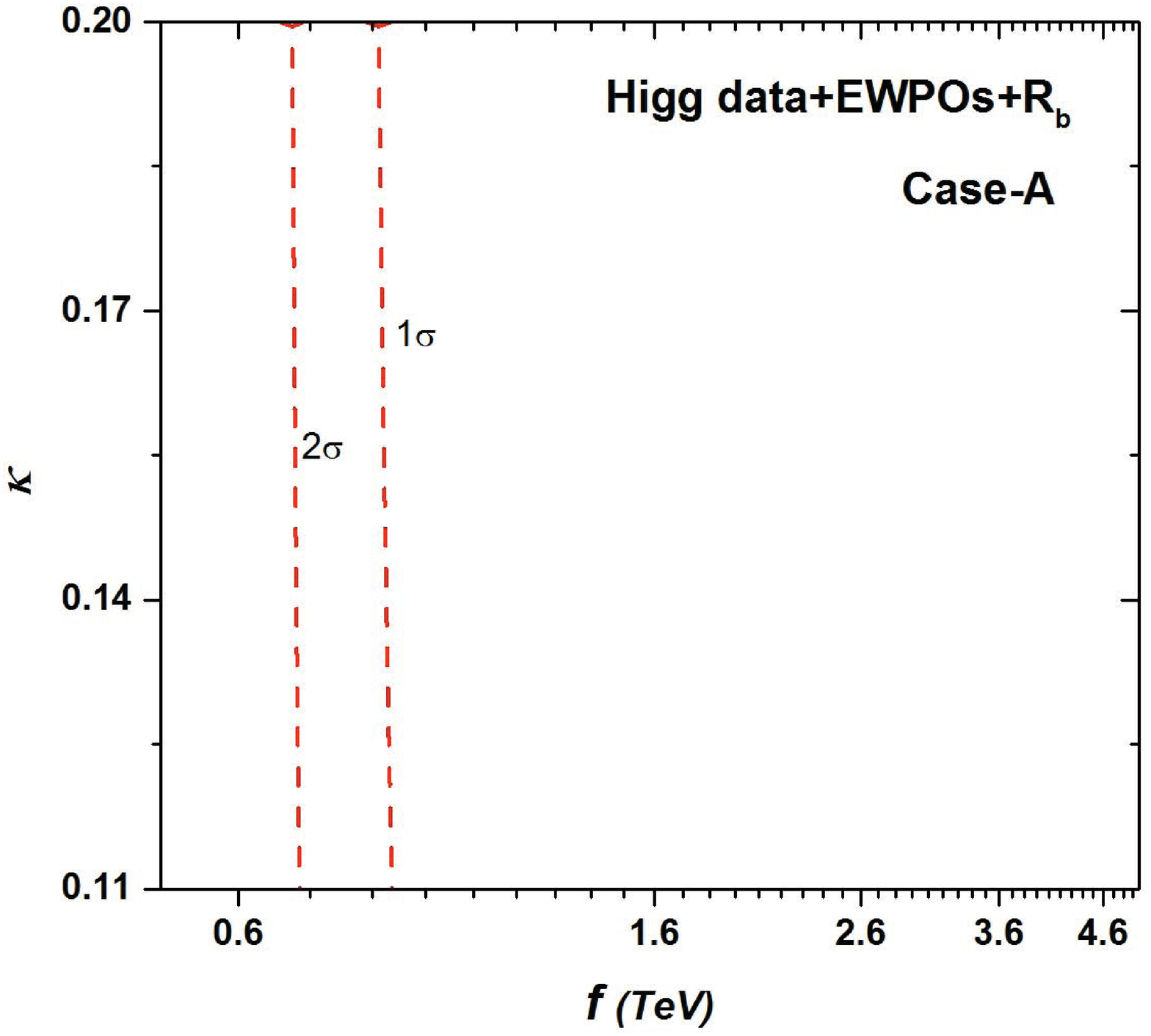}
\includegraphics[width=7cm,height=8cm]{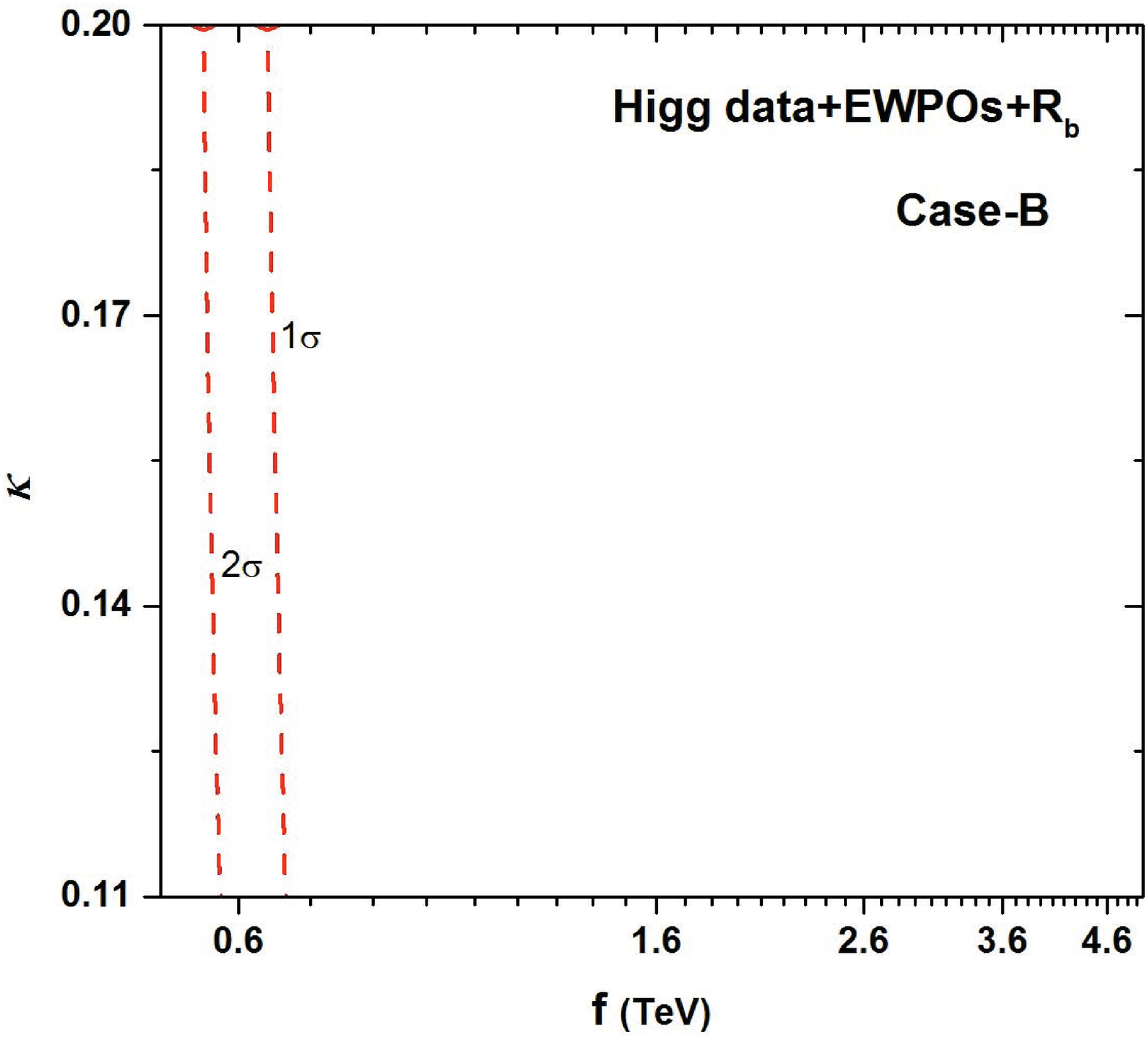}
\caption{Excluded regions (left each contour) in the plane of $\kappa$ versus $f$ for Case A and Case B, where the parameter $R$ is marginalized over.}
\label{fig:HiggsEWPO}
\end{figure}

We scan over the free parameters $\kappa$, $f$ and $R$ within the following ranges,
\begin{eqnarray}
500 ~{\rm GeV} < f < 5000 ~{\rm GeV}, \quad 0.11 < \kappa < 0.2, \quad 0.1 < R < 3.3.
\end{eqnarray}
where we assume $\kappa_\ell=\kappa_q=\kappa$. In order to escape LHC limits from the multijet with $\met$, we require $\kappa \leqslant 0.2$ to forbidden $T$-odd fermions decay to the heavy gauge bosons $Z_H$ and $W_H$. Besides, we decouple the $T$-odd top quark $t_-$ by setting $m_{t_-}=3$ TeV in order to avoid the bound of LHC searches for long-lived charged particles. We adopt our previous scan method \cite{lht-fit-yang-1,lht-fit-yang-2} by constructing the likelihood $\cal{L}\equiv$exp$[-\displaystyle{\sum} \chi_{i}^{2}]$ for each point, where index $i$ denotes the following constraint:
\begin{itemize}
\item[(1)] The electroweak precision observables: $S$, $T$ and $U$ \cite{lht-ewpos-1}. We use the experimental values of $S$, $T$ and $U$ from Ref. \cite{pdg}.

\item[(2)] $R_{b}$ \cite{lht-rb-2}. We use the final combined result $R_{b} = 0.21629\pm 0.00066$ \cite{pdg} from the LEP and SLD measurements.

\item[(3)] Higgs data. We check the LHT Higgs couplings by using \textsf{HiggsSignals-1.4.0} \cite{higgssignals}, which includes the available Higgs data sets from the ATLAS, CMS, CDF and D0 collaborations. The mass-centered $\chi^2$ method is chosen in our study.
\end{itemize}
On the other hand, since the SM flavor symmetry is broken by the extension of the top quark sector, the mixing between top partner and down-type quark can induce flavor changing neutral current processes at one-loop level \cite{lht-flavor-1,lht-flavor-2,lht-flavor-3,lht-flavor-4}. We checked our samples and found that the constraints from $B_s \to \mu^+\mu^-$ \cite{bs-exp-1,bs-exp-2} can be easily satisfied within the current uncertainty.

In Fig.~\ref{fig:HiggsEWPO}, we show the excluded regions (left each contour) in the plane of $\kappa$ versus $f$ for Case A and Case B by fitting Higgs data, EWPOs and $R_b$. The parameter $R$ is marginalized over. From the Fig.~\ref{fig:HiggsEWPO}, it can be seen that the lower bound on the symmetry breaking scale $f$ is almost independent of $\kappa$ and has reached about 675 (550) GeV at $2\sigma$ level in Case A (B), which corresponds to $m_{A_H}=99 (76)$ GeV. Since the reduced bottom Yukawa coupling in Case B is smaller than that in Case A (cf. Eq.~(\ref{hdd})), the non-fermionic decays of the Higgs boson can be  enhanced in Case B, which is more consistent with the current ATLAS-8 TeV Higgs data. So the lower bound on $f$ in Case B is weaker than that in Case A. To conservatively examine dark matter and LHC experiment constraints on heavy photon $A_H$, we will focus on Case A in the following.

\subsection{Planck Relic Abundance and PandaX-II/LUX 2016 Direct Detections}

\begin{figure}[ht]
\centering
\includegraphics[width=8cm,height=9cm]{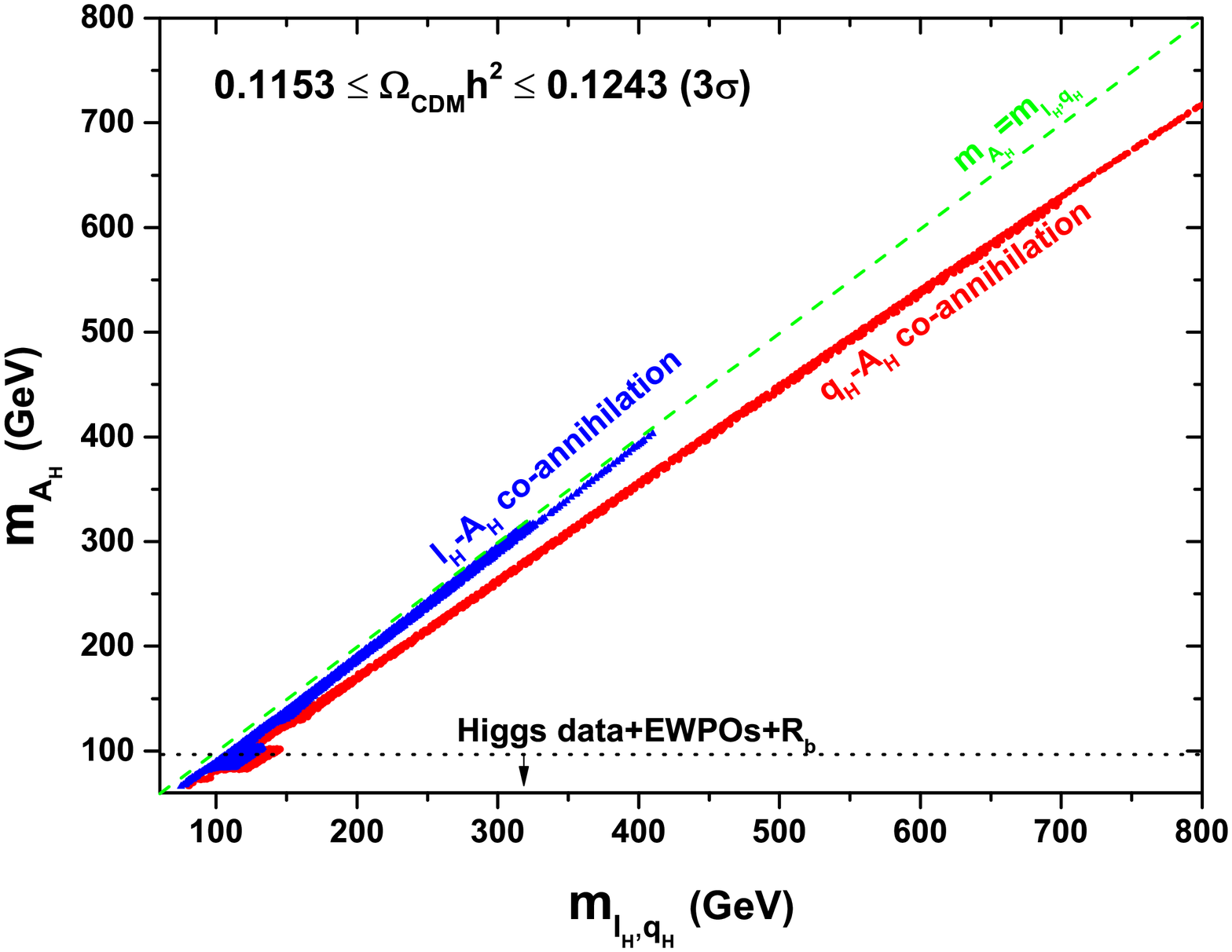}
\hspace{-1.5cm}
\includegraphics[width=8cm,height=9cm]{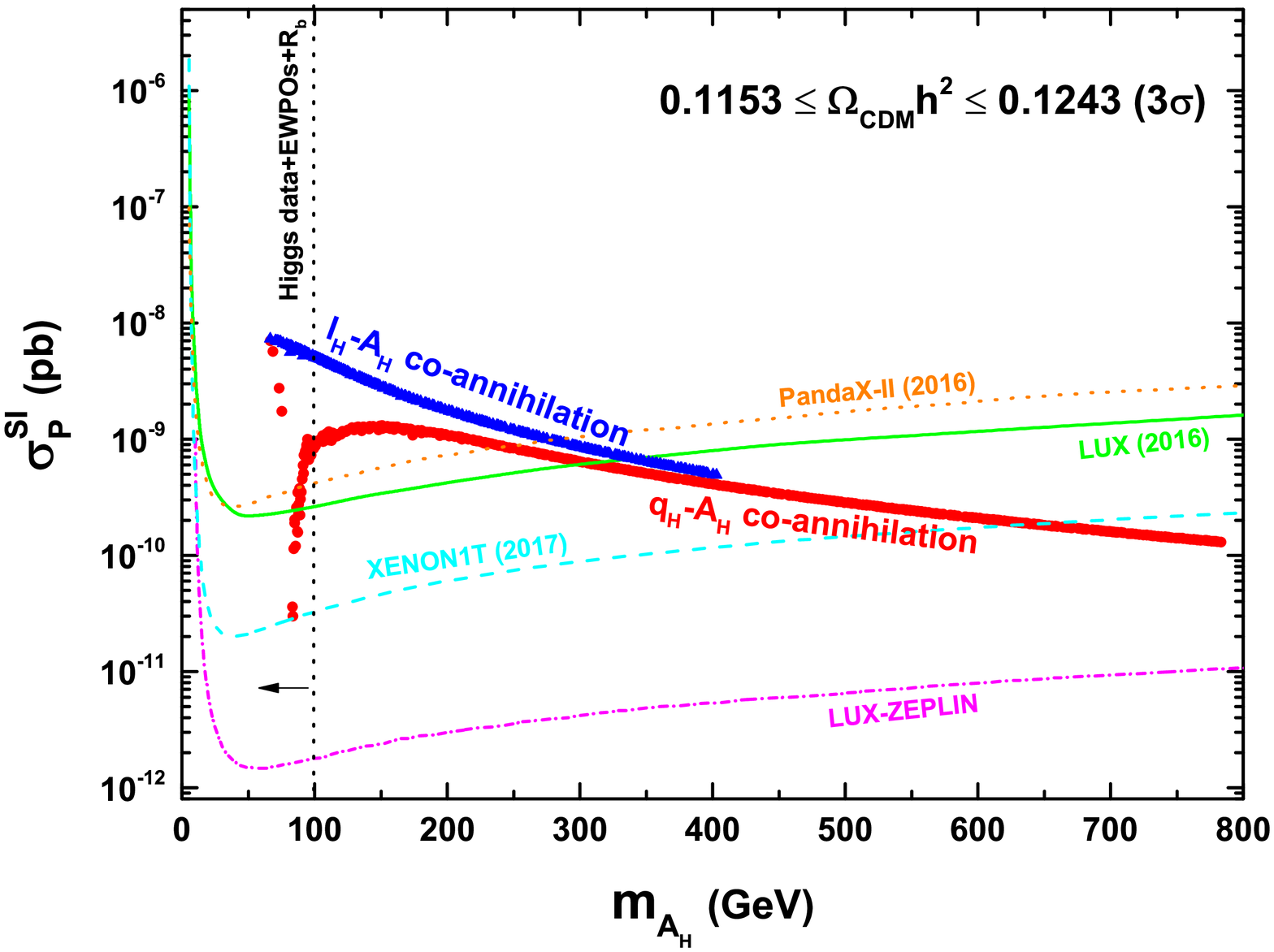}
\caption{The scatter plot on the planes of $m_{A_H}$ versus $m_{\ell_H,q_H}$ (left panel) and $\sigma^{SI}_p$ versus $m_{A_H}$ (right panel). All samples satisfy the Planck dark matter relic abundance within $3\sigma$ range. For $\ell_H-A_H$ ($q_H-A_H$) co-annihilation, $\kappa_q$ ($\kappa_\ell$) is fixed at 3.}
\label{fig:omega}
\end{figure}
In the LHT, $A_H$ pair mainly annihilates through a Higgs boson in $s$-channel to a pair of fermions, $W/Z$ bosons and Higgs bosons, whose contributions to the relic density depend on the mass of $A_H$. When $m_{A_H}$ is heavier than $m_h/2$, the Higgs resonance effect in $A_H$ pair annihilation will become small and the $A_H A_H$ annihilation cross section will be too small to give the right relic abundance. We use the \textsf{MicrOMEGAs4.2.5} \cite{Belanger:2014vza} to calculate the relic density $\Omega h^2$ and the spin-independent scattering cross section between DM and nucleon $\sigma^{SI}_p$.

In the left panel of Fig.~\ref{fig:omega}, we show the scatter plot on the plane of $m_{A_H}$ versus $m_{\ell_H,q_H}$. We require samples to satisfy the Planck dark matter relic abundance within $3\sigma$ range. We can see that the constraint of the relic density requires $A_H$ co-annihilate with $T$-odd fermions, which is typically given by,
\begin{eqnarray}
\frac{\Delta m_{\ell_H}}{m_{A_H}}=\frac{m_{\ell_H} - m_{A_H}}{m_{A_H}}\lesssim 0.1\\
\frac{\Delta m_{q_H}}{m_{A_H}}=\frac{m_{q_H} - m_{A_H}}{m_{A_H}}\lesssim 0.2
\label{coannihilation}
\end{eqnarray}
In the calculation of co-annihilation, the effective dark matter annihilation cross section $\sigma_{eff}(A_H)$ includes the contributions from $A_H$ pair annihilation, $A_H$ and $\ell_H/q_H$ co-annihilation and $\ell_H/q_H$ pair annihilation \cite{co-dm}. For the colored co-annihilation partner $q_H$, the contribution of $q_H$ pair annihilation is large because of the strong coupling. While for the non-colored co-annihilation partner $\ell_H$, three contributions are comparable. So the annihilation cross section of $A_H-\ell_H$ is smaller than that of $A_H-q_H$ for the given mass splitting \cite{Baker:2015qna}. To obtain the correct relic density, the mass splitting between $A_H$ and co-annihilation partner in $A_H-\ell_H$ co-annihilation has to be smaller than that in $A_H-\ell_q$ co-annihilation (c.f. Eq.~\ref{coannihilation}). When $A_H$ becomes heavy, the effective cross section $\sigma_{eff}(A_H)$ decreases so that the dark matter relic density will be overproduced in the universe. This leads to the upper bounds on the masses of dark matter and its co-annihilation partners. Due to the small co-annihilation cross section, the resulting viable region of parameter space for the dark matter relic density only extends to about 400 GeV in $A_H-\ell_H$ co-annihilation.

In the right panel of Fig.~\ref{fig:omega}, we show the scatter plot on the plane of $\sigma^{SI}_p$ versus $m_{A_H}$. There are three processes contributing to the cross-section of $A_H$ scattering off nucleon: Higgs-boson-exchanged $t$-channel, $T$-odd-quark-exchanged $t$-channel and $s$-channel \cite{lht-dm}. For $\ell_H-A_H$ co-annihilation, the dominant contribution to $\sigma^{SI}_p$ is the Higgs-boson-exchanged $t$-channel since the $T$-odd quarks are decoupled. The mass of $A_H$ can be excluded up to about 380 (270) GeV by the LUX (PandaX-II) 2016 data \cite{lux,pandaX}. While for $q_H-A_H$ co-annihilation, $m_{A_H}<350 (240)$ GeV is excluded by the LUX (PandaX-II) 2016 data. This is because that the cancellation between $T$-odd quark and the top quark loops in $hgg$ coupling reduces the contribution of Higgs-boson-exchanged $t$-channel to cross section $\sigma^{SI}_p$. Besides, the amplitudes of $T$-odd-quark-exchanged $t$-channel and $s$-channel interference destructively in our parameter space. The expected XENON1T (2017) experiment \cite{xenon1t} will allow it to cover $\ell_H-A_H$ co-annihilation region and push the lower limit of $m_{A_H}$ up to 640 GeV.

\subsection{ATLAS-8 TeV Monojet limit}

\begin{figure}[ht]
\includegraphics[width=16cm,height=8cm]{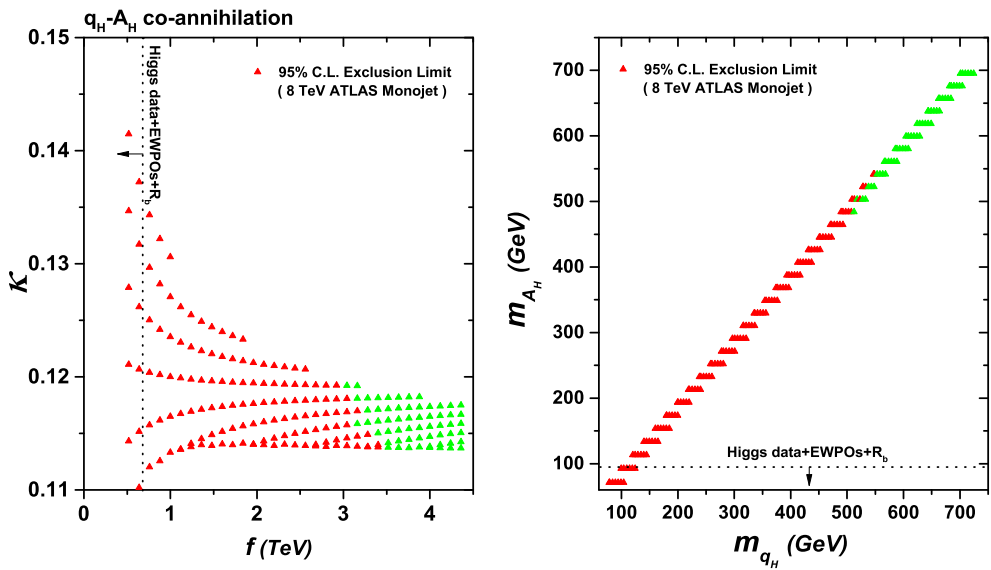}
\caption{Same as Fig.~\ref{fig:omega}, but for ATLAS-8 TeV monojet constraint on $q_H$-$A_H$ co-annihilation.}
\label{fig:monojet}
\end{figure}
In co-annihilations, the decay products of light $T$-odd lepton or quark are usually very soft. One way of probing such a compressed region is to use the ISR/FSI jet to boost the soft objects, which produces the monojet(-like) events at the LHC \cite{wu-1,wu-2,wu-3}. Since the $T$-odd leptons are produced via the electroweak interaction, the cross section of the heavy lepton pair production are much smaller than that of the heavy quark pair production at the LHC. We checked and found that the ATLAS monojet data can not give an exclusion limit on the $\ell_H-A_H$ co-annihilation scenario. So we only present the results for $q_H-A_H$ co-annihilation in our work.

We recast the ATLAS-8 TeV monojet bound \cite{monojet} by using \textsf{CheckMATE-1.2.1} \cite{checkmate}. In our scenario, the monojet events arise from the processes:
\begin{eqnarray}
pp \to j q_H A_H, \quad j q_H \bar{q}_H
\end{eqnarray}
We generate the parton level signal events by using \textsf{MadGraph5\_aMC@NLO} \cite{mad5}. Then, the parton level events are showered and hadronized by \textsf{PYTHIA} \cite{pythia}. The fast detector simulation are performed with the tuned \textsf{Delphes} \cite{delphes}. The jet is clustered by \textsf{FastJet} \cite{fastjet} with the anti-$k_t$ algorithm \cite{anti-kt}. We normalize the cross section of $q_H \bar{q}_H$ and $q_H A_H$ productions to their NLO value by including a $K$-factor $1.5$ \cite{k-factor}. Finally, we define the ratio $r = max(N_{S,i}/S^{95\%}_{obs,i})$ to estimate the exclusion limit. Here $N_{S,i}$ is the event number of signal for $i$-th signal region and $S^{95\%}_{obs,i}$ is the corresponding observed 95\% C.L. upper limit. The max is over all signal regions in the analysis. We conclude that a point is excluded at 95\% C.L. if $r > 1$. In Fig.~\ref{fig:monojet}, we show the monojet constraints on the parameter space of $q_H-A_H$ co-annihilation. We can see that the monojet limit can exclude the scale $f$ up to 3.4 TeV, which corresponds to $m_{A_H}>540$ GeV. For a given $f$, the monojet has a better sensitivity in the region with small Yukawa coupling $\kappa_q$.

\section{Conclusions}\label{section4}
In this work, we investigate the lower limit on the mass of $A_H$ dark matter by using the constraints from Higgs data, EWPOs, $R_b$, Planck dark matter relic abundance, LUX direct detection and LHC-8 TeV monojet results. We find that the mass of $A_H$ has been excluded up to 99 GeV by Higgs data, EWPOs and $R_b$. Therefore, $A_H$ needs to co-annihilate with $T$-odd quarks ($q_H$) or leptons ($\ell_H$) to give the correct dark matter relic abundance. Further, with the very recent LUX (PandaX-II) 2016 data, the lower limit of $m_{A_H}$ can be pushed up to about 380 (270) GeV and 350 (240) GeV for $\ell_H$-$A_H$ and $q_H-A_H$ co-annihilations, respectively. Also, we find that ATLAS 8 TeV monojet result can give a stringent lower limit, $m_{A_H}>540$ GeV, for $q_H$-$A_H$ co-annihilation, while can not produce the limit on $m_{A_H}$ for $l_H-A_H$ co-annihilation. In future XENON1T (2017) experiment, parameter space of $\ell_H$-$A_H$ co-annihilation can be fully covered and the lower limit of $m_{A_H}$ will be pushed up to about 640 GeV for $q_H$-$A_H$ co-annihilation.

\acknowledgments
Lei Wu thanks the helpful discussions with Dr. Lei Wang. This work is partly supported by the Australian Research Council, by the National Natural Science Foundation of China (NNSFC) under grants Nos. 11275057, 11305049, 11375001 and 11405047, by Specialised Research Fund for the Doctoral Program of Higher Education under Grant No. 20134104120002 and by the Startup Foundation for Doctors of Henan Normal University under contract No.11112, by the China Postdoctoral Science Foundation under Grant No. 2014M561987 and the Joint Funds of the National Natural Science Foundation of China (U1404113).




\begin{thebibliography}{99}

\bibitem{higgs-atlas}
  G.~Aad {\it et al.}  [ATLAS Collaboration],
  Phys.\ Lett.\ B {\bf 716}, 1 (2012).

\bibitem{higgs-cms}
  S.~Chatrchyan {\it et al.}  [CMS Collaboration],
  Phys.\ Lett.\ B {\bf 716}, 30 (2012).

\bibitem{lht-1}
  H.~-C.~Cheng and I.~Low,
  JHEP {\bf 0309}, 051 (2003);

\bibitem{lht-2}
  H.~-C.~Cheng and I.~Low,
  JHEP {\bf 0408}, 061 (2004);

\bibitem{lht-3}
  I.~Low,
  JHEP {\bf 0410}, 067 (2004);

\bibitem{lh-1}
  N.~Arkani-Hamed, A.~G.~Cohen and H.~Georgi,
  Phys.\ Lett.\ B {\bf 513}, 232 (2001);

\bibitem{lh-2}
  N.~Arkani-Hamed, A.~G.~Cohen, E.~Katz, A.~E.~Nelson, T.~Gregoire and J.~G.~Wacker,
  JHEP {\bf 0208}, 021 (2002);

\bibitem{lh-3}
  N.~Arkani-Hamed, A.~G.~Cohen, E.~Katz and A.~E.~Nelson,
  JHEP {\bf 0207}, 034 (2002).


\bibitem{lh-ewpos-1}
  C.~Csaki, J.~Hubisz, G.~D.~Kribs, P.~Meade and J.~Terning,
  Phys.\ Rev.\ D {\bf 67}, 115002 (2003);

\bibitem{lh-ewpos-2}
  C.~Csaki, J.~Hubisz, G.~D.~Kribs, P.~Meade and J.~Terning,
  Phys.\ Rev.\ D {\bf 68}, 035009 (2003);

\bibitem{lh-ewpos-3}
  J.~L.~Hewett, F.~J.~Petriello and T.~G.~Rizzo,
  JHEP {\bf 0310}, 062 (2003);

\bibitem{lh-ewpos-4}
  M.~C.~Chen and S.~Dawson,
  Phys.\ Rev.\ D {\bf 70}, 015003 (2004);

\bibitem{lh-ewpos-5}
  W.~Kilian and J.~Reuter,
  Phys.\ Rev.\ D {\bf 70}, 015004 (2004);

\bibitem{lh-ewpos-6}
  G.~Marandella, C.~Schappacher and A.~Strumia,
  Phys.\ Rev.\ D {\bf 72}, 035014 (2005).


\bibitem{lht-ewpos-1}
  J.~Hubisz, P.~Meade, A.~Noble and M.~Perelstein,
  JHEP {\bf 0601}, 135 (2006).

\bibitem{lht-phe-1}
  T.~Han, H.~E.~Logan, B.~McElrath and L.~T.~Wang,
  Phys.\ Rev.\ D {\bf 67}, 095004 (2003).

\bibitem{lht-phe-2}
  J.~Hubisz and P.~Meade,
  Phys.\ Rev.\ D {\bf 71}, 035016 (2005);

\bibitem{lht-phe-3}
  C.~R.~Chen, K.~Tobe and C.-P.~Yuan,
  Phys.\ Lett.\ B {\bf 640}, 263 (2006).



\bibitem{lht-phe-4}
  H.~C.~Cheng, I.~Low and L.~T.~Wang,
  Phys.\ Rev.\ D {\bf 74}, 055001 (2006);

\bibitem{lht-phe-4.5}
  A.~Freitas and D.~Wyler,
  JHEP {\bf 0611}, 061 (2006)
  doi:10.1088/1126-6708/2006/11/061
  [hep-ph/0609103].

\bibitem{lht-phe-5}
  A.~Belyaev, C.~R.~Chen, K.~Tobe and C.-P.~Yuan,
  Phys.\ Rev.\ D {\bf 74}, 115020 (2006);

\bibitem{lht-phe-5.5}
  D.~Choudhury and D.~K.~Ghosh,
  JHEP {\bf 0708}, 084 (2007).

\bibitem{lht-phe-6}
  Q.~H.~Cao and C.~R.~Chen,
  Phys.\ Rev.\ D {\bf 76}, 075007 (2007).

\bibitem{lht-phe-7}
  S.~Matsumoto, M.~M.~Nojiri and D.~Nomura,
  Phys.\ Rev.\ D {\bf 75}, 055006 (2007);

\bibitem{lht-phe-8}
  Q.~H.~Cao, C.~S.~Li and C.-P.~Yuan,
  Phys.\ Lett.\ B {\bf 668}, 24 (2008);

\bibitem{lht-phe-9}
  L.~Wang and J.~M.~Yang,
  Phys.\ Rev.\ D {\bf 77}, 015020 (2008).


\bibitem{lht-phe-10}
  Q.~H.~Cao, C.~R.~Chen, F.~Larios and C.-P.~Yuan,
  Phys.\ Rev.\ D {\bf 79}, 015004 (2009).

\bibitem{lht-phe-11}
  J.~Y.~Liu, Z.~G.~Si and C.~X.~Yue,
  Phys.\ Rev.\ D {\bf 81}, 015011 (2010).

\bibitem{lht-phe-12}
  S.~Matsumoto, T.~Moroi and K.~Tobe,
  Phys.\ Rev.\ D {\bf 78}, 055018 (2008);

\bibitem{lht-phe-13}
  S.~Yang,
  Phys.\ Lett.\ B {\bf 675}, 352 (2009);

\bibitem{lht-phe-14}
  D.~Choudhury, D.~K.~Ghosh and S.~K.~Rai,
  JHEP {\bf 1207}, 013 (2012).

\bibitem{lht-phe-15}
  R.~Y.~Zhang, {\it et al.},
  Phys.\ Rev.\ D {\bf 85}, 015017 (2012);

\bibitem{lht-phe-16}
  S.~J.~Xiong, {\it et al.},
  Phys.\ Rev.\ D {\bf 89}, no. 11, 114015 (2014);

\bibitem{lht-phe-16.5}
  L.~W.~Chen, R.~Y.~Zhang, W.~G.~Ma, W.~H.~Li, P.~F.~Duan and L.~Guo,
  Phys.\ Rev.\ D {\bf 90}, no. 5, 054020 (2014).

\bibitem{lht-phe-17}
  B.~Yang, J.~Han and N.~Liu,
  JHEP {\bf 1504}, 148 (2015);


\bibitem{lht-phe-18}
  N.~Liu, L.~Wu, B.~Yang and M.~Zhang,
  Phys.\ Lett.\ B {\bf 753}, 664 (2016).

\bibitem{lht-phe-19}
  Q.~H.~Cao, C.~R.~Chen and Y.~Liu,
  arXiv:1512.09144 [hep-ph].

\bibitem{lht-phe-20}
  D.~Choudhury, D.~K.~Ghosh, S.~K.~Rai and I.~Saha,
  JHEP {\bf 1606}, 074 (2016).


\bibitem{Servant:2002hb}
  G.~Servant and T.~M.~P.~Tait,
  New J.\ Phys.\  {\bf 4}, 99 (2002)
  doi:10.1088/1367-2630/4/1/399
  [hep-ph/0209262].

\bibitem{lht-dm}
  A.~Birkedal, A.~Noble, M.~Perelstein and A.~Spray,
  Phys.\ Rev.\ D {\bf 74}, 035002 (2006).


\bibitem{lht-dm-wanglei}
  L.~Wang, J.~M.~Yang and J.~Zhu,
  Phys.\ Rev.\ D {\bf 88}, no. 7, 075018 (2013).


\bibitem{lht-dm-chuanren}
  C.~R.~Chen, M.~C.~Lee and H.~C.~Tsai,
  JHEP {\bf 1406}, 074 (2014).



\bibitem{lht-fit-tonini-1}
  J.~Reuter, M.~Tonini and M.~de Vries,
  JHEP {\bf 1402}, 053 (2014).

\bibitem{lht-fit-tonini-2}
  J.~Reuter and M.~Tonini,
  JHEP {\bf 1302}, 077 (2013).

\bibitem{lht-fit-yang-1}
  B.~Yang, G.~Mi and N.~Liu,
  JHEP {\bf 1410}, 47 (2014).

\bibitem{lht-fit-yang-2}
  C.~Han, A.~Kobakhidze, N.~Liu, L.~Wu and B.~Yang,
  Nucl.\ Phys.\ B {\bf 890}, 388 (2014).

\bibitem{lht-fit-wang}
  X.~F.~Han, L.~Wang, J.~M.~Yang and J.~Zhu,
  Phys.\ Rev.\ D {\bf 87}, no. 5, 055004 (2013).

\bibitem{Baker:2015qna}
  M.~J.~Baker {\it et al.},
  JHEP {\bf 1512}, 120 (2015)
  doi:10.1007/JHEP12(2015)120
  [arXiv:1510.03434 [hep-ph]].

\bibitem{zyhan}
  Z.~Han and W.~Skiba,
  Phys.\ Rev.\ D {\bf 71}, 075009 (2005)
  [hep-ph/0412166].

\bibitem{pdg}
  K.~A.~Olive {\it et al.} [Particle Data Group Collaboration],
  Chin.\ Phys.\ C {\bf 38}, 090001 (2014).

\bibitem{lht-rb-1}
  C.~x.~Yue and W.~Wang,
  Nucl.\ Phys.\ B {\bf 683}, 48 (2004);

\bibitem{lht-rb-2}
  X.~F.~Han,
  Phys.\ Rev.\ D {\bf 80}, 055027 (2009);


\bibitem{lht-rb-3}
  B.~Yang, X.~Wang and J.~Han,
  Nucl.\ Phys.\ B {\bf 847}, 1 (2011).

\bibitem{tp-atlas}
  G.~Aad {\it et al.} [ATLAS Collaboration],
  JHEP {\bf 1411}, 104 (2014)
  [arXiv:1409.5500 [hep-ex]].
  The ATLAS collaboration [ATLAS Collaboration],
  ATLAS-CONF-2015-012, ATLAS-COM-CONF-2015-012.

\bibitem{tp-cms}
  S.~Chatrchyan {\it et al.} [CMS Collaboration],
  Phys.\ Lett.\ B {\bf 729}, 149 (2014)
  [arXiv:1311.7667 [hep-ex]].

\bibitem{higgssignals}
  P.~Bechtle {\it et al.},
  Eur.\ Phys.\ J.\ C {\bf 74}, 2711 (2014)  [arXiv:1305.1933 [hep-ph]];
  P.~Bechtle {\it et al.},
  Comput.\ Phys.\ Commun.\  {\bf 181}, 138 (2010)  [arXiv:0811.4169 [hep-ph]].

\bibitem{lht-flavor-1}
  J.~Hubisz, S.~J.~Lee and G.~Paz,
  JHEP {\bf 0606}, 041 (2006).

\bibitem{lht-flavor-2}
  M.~Blanke, A.~J.~Buras, A.~Poschenrieder, C.~Tarantino, S.~Uhlig and A.~Weiler,
  JHEP {\bf 0612}, 003 (2006).

\bibitem{lht-flavor-3}
  M.~Blanke, A.~J.~Buras, A.~Poschenrieder, S.~Recksiegel, C.~Tarantino, S.~Uhlig and A.~Weiler,
  JHEP {\bf 0701}, 066 (2007).

\bibitem{lht-flavor-4}
  M.~Blanke, A.~J.~Buras and S.~Recksiegel,
  arXiv:1507.06316 [hep-ph].

\bibitem{bs-exp-1}
[CMS Collaboration] CMS-PAS-BPH-13-007;

\bibitem{bs-exp-2}
[LHCb Collaboration] LHCb-CONF-2013-012;


\bibitem{Belanger:2014vza}
  G.~B¨¦langer, F.~Boudjema, A.~Pukhov and A.~Semenov,
  Comput.\ Phys.\ Commun.\  {\bf 192}, 322 (2015).

\bibitem{co-dm}
  K.~Griest and D.~Seckel,
  Phys.\ Rev.\ D {\bf 43}, 3191 (1991).
  doi:10.1103/PhysRevD.43.3191

\bibitem{lux}
  D.~S.~Akerib {\it et al.},
  arXiv:1608.07648 [astro-ph.CO].

\bibitem{pandaX}
  A.~Tan {\it et al.} [PandaX-II Collaboration],
  Phys.\ Rev.\ Lett.\  {\bf 117}, no. 12, 121303 (2016)
  doi:10.1103/PhysRevLett.117.121303
  [arXiv:1607.07400 [hep-ex]].

\bibitem{xenon1t}
  E.~Aprile {\it et al.} [XENON Collaboration],
  JCAP {\bf 1604}, no. 04, 027 (2016)
  doi:10.1088/1475-7516/2016/04/027
  [arXiv:1512.07501 [physics.ins-det]].


\bibitem{wu-1}
  C.~Han, A.~Kobakhidze, N.~Liu, A.~Saavedra, L.~Wu and J.~M.~Yang,
  JHEP {\bf 1402}, 049 (2014).

\bibitem{wu-2}
  C.~Han, L.~Wu, J.~M.~Yang, M.~Zhang and Y.~Zhang,
  Phys.\ Rev.\ D {\bf 91}, 055030 (2015).

\bibitem{wu-3}
  K.~i.~Hikasa, J.~Li, L.~Wu and J.~M.~Yang,
  Phys.\ Rev.\ D {\bf 93}, no. 3, 035003 (2016).


\bibitem{monojet}
  G.~Aad {\it et al.} [ATLAS Collaboration],
  Eur.\ Phys.\ J.\ C {\bf 75}, no. 7, 299 (2015)
  Erratum: [Eur.\ Phys.\ J.\ C {\bf 75}, no. 9, 408 (2015)].

\bibitem{checkmate}
  M.~Drees {\it et al.},
  Comput.\ Phys.\ Commun.\  {\bf 187}, 227 (2014).
  J.~S.~Kim {\it et al.},
  arXiv:1503.01123 [hep-ph].


\bibitem{mad5}
  J.~Alwall {\it et al.},
  JHEP {\bf 1407}, 079 (2014).

\bibitem{pythia}
  T.~Sjostrand, S.~Mrenna and P.~Z.~Skands,
  JHEP {\bf 0605}, 026 (2006).

\bibitem{delphes}
  J.~de Favereau,  {\it et al.},
arXiv:1307.6346 [hep-ex].

\bibitem{fastjet}
  M.~Cacciari, G.~P.~Salam and G.~Soyez,
  Eur.\ Phys.\ J.\ C {\bf 72}, 1896 (2012)
  [arXiv:1111.6097 [hep-ph]].

\bibitem{anti-kt}
  M.~Cacciari, G.~P.~Salam and G.~Soyez,
  JHEP {\bf 0804}, 063 (2008).

\bibitem{k-factor}
  Y.~Xiao-Dong, X.~Shou-Jian, M.~Wen-Gan, Z.~Ren-You, G.~Lei and L.~Xiao-Zhou,
  Phys.\ Rev.\ D {\bf 89}, no. 1, 014008 (2014).

\end{thebibliography}


\end{document}